\documentclass[aps,prb,twocolumn,superscriptaddress,showpacs]{revtex4}
\usepackage{graphicx,bm,color,float,textcomp,mathtools}

\newcommand\seo{$\rm SrEr_2O_4$}
\newcommand\sdo{$\rm SrDy_2O_4$}
\newcommand\sho{$\rm SrHo_2O_4$}
\newcommand\syo{$\rm SrYb_2O_4$}
\newcommand\slo{Sr$Ln_2$O$_4$}

\begin{document}

\title{Highly frustrated magnetism in SrHo$_2$O$_4$: Coexistence of two types of short-range order}

\date{\today}

\author{O. Young}
\affiliation{Department of Physics, University of Warwick, Coventry CV4 7AL, United Kingdom}
\author{A. R. Wildes}
\affiliation{Institut Laue-Langevin, 6 Jules Horowitz, BP156, 38042 Grenoble Cedex 9, France}
\author{P. Manuel}
\affiliation{ISIS Facility, Rutherford Appleton Laboratory, Chilton, Didcot OX11 0QX, United Kingdom}
\author{B. Ouladdiaf}
\affiliation{Institut Laue-Langevin, 6 Jules Horowitz, BP156, 38042 Grenoble Cedex 9, France}
\author{D. D. Khalyavin}
\affiliation{ISIS Facility, Rutherford Appleton Laboratory, Chilton, Didcot OX11 0QX, United Kingdom}
\author{G. Balakrishnan}
\affiliation{Department of Physics, University of Warwick, Coventry CV4 7AL, United Kingdom}
\author{O. A. Petrenko}
\affiliation{Department of Physics, University of Warwick, Coventry CV4 7AL, United Kingdom}

\begin{abstract}
\sho\ is a geometrically frustrated magnet in which the magnetic Ho$^{3+}$ ions form honeycomb layers connected through a network of zigzag chains. 
At low-temperature two distinct types of short-range magnetic order can be inferred from single crystal diffraction data, collected using both polarized and unpolarized neutrons.
In the ($hk0$) plane the diffuse scattering is most noticeable around the {\bf k}~=~0 positions and its intensity rapidly increases at temperatures below 0.7~K. 
In addition, planes of diffuse scattering at $Q$~=~($hk\pm\!\!\frac{l}{2}$) are visible at temperatures as high as 4.5~K.
These planes coexist with the broad peaks of diffuse scattering in the ($hk0$) plane at low-temperatures.
Correlation lengths associated with the broad peaks are $L$~$\approx$~150~\AA\ in the \emph{a-b} plane and $L$~$\approx$~190~\AA\ along $c$ axis, while the correlation length associated with the diffuse scattering planes is $L$~$\approx$~230~\AA\ along the $c$ axis at the lowest temperature.
Both types of diffuse scattering are elastic in nature.
The highly unusual coexistance of the two types of diffuse scattering in \sho\ is likely to be the result of the presence of two crystallographically inequivalent sites for Ho$^{3+}$ in the unit cell.
\end{abstract}

\pacs{75.25.-j, 75.50.Ee, 75.47.Lx}


\maketitle

\section{Introduction}

The term magnetic frustration describes the inability of a magnetic system to satisfy all the competing interactions in order to establish a unique ground state.~\cite{Diep_2005, Buschow_2001, Ramirez_1994}
In the presence of antiferromagnetic exchange interactions, magnets based on corner- or edge-sharing triangles~\cite{Collins_1997} or tetrahedra~\cite{Gardner_2010} can exhibit geometric frustration.
Common examples of systems in which frustration plays a fundamental role in establishing highly degenerate ground states and low-temperature properties such as reduced critical temperatures include the kagome,~\cite{Ramirez_1990, Chalker_1992} pyrochlore~\cite{Bramwell_2001, Gardner_2010} and garnet~\cite{Schiffer_1994, Yoshioka_2004} structures.
New magnets which exhibit the effects of geometric frustration are constantly being discovered. 
The behaviour of these systems is often complex, with many magnets remaining disordered down to very low-temperatures - traditional examples being spin liquids~\cite{Moessner_1998, Canals_1998} and spin glasses.~\cite{Villain_1979, Binder_1986}

\begin{figure}[tb]
\centering
\includegraphics[width=0.60\columnwidth]{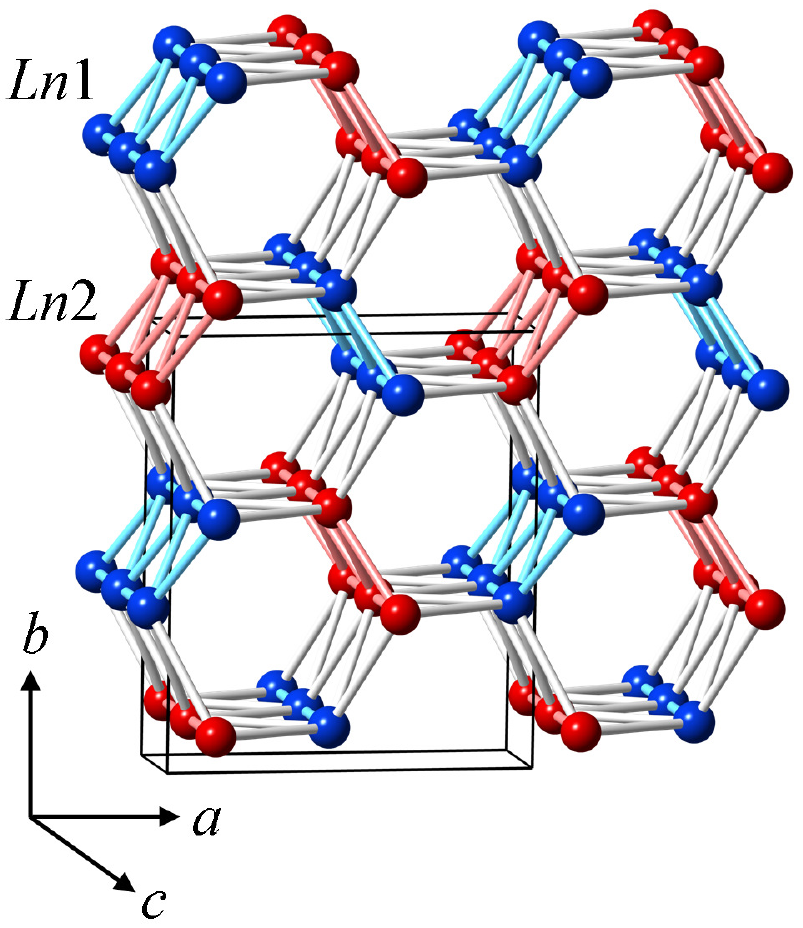}
\caption{(Color online) Magnetic sublattice of \slo\ compounds, with the two crystallographically inequivalent positions of the rare earth ions shown in different colors. 
When viewed in the \emph{a-b} plane, honeycombs of the $Ln^{3+}$ ions are visible. 
Zigzag chains running along the $c$ axis connect the honeycomb layers and give rise to geometric frustration.
The box indicates the dimensions of the unit cell.}
\label{Lnsublattice}
\end{figure}

In this paper we report on an investigation of single crystal samples of \sho, a compound that displays no long-range order down to 0.05~K. 
More intriguingly, there appears to be two distinct types of short-range magnetic order coexisting in this material, which has no chemical instability or phase separation.
\sho\ belongs to the \slo\ (where \emph{Ln} = Lanthanide) family of compounds,~\cite{Lopato} which crystallise in the form of calcium ferrite,~\cite{Decker_1957} space group \emph{Pnam}.
This allows the magnetic \emph{Ln} ions to be linked in a network of triangles and hexagons, as shown in Fig.~\ref{Lnsublattice}, with two crystallographically inequivalent sites of the rare earth ions shown in red and blue. 
We believe that it is likely that the two different types of short-range order observed in \sho\ can be attributed to the two different sites for the Ho$^{3+}$ ions in the unit cell.

For all the \slo\ systems the magnetic ions form zigzag chains that run along the $c$ axis.
These triangular ladders can be frustrated, and are magnetically equivalent to one-dimensional chains with first- and second-nearest-neighbour interactions. 
The chains of $Ln^{3+}$ ions interconnect by forming a distorted \emph{honeycomb} structure, a bipartite lattice made up of edge sharing hexagons, in the \emph{a-b} plane.
The honeycomb lattice would only be frustrated if further neighbor exchange is considered, but receives substantial theoretical interest since it has the smallest possible coordination number in two-dimensions.~\cite{Mattsson_1994, Okumura_2010, Yu_2013}
Thus for these systems quantum fluctuations are expected to have a much larger effect on the stability of magnetic order than, for example, for square or triangular lattices. 
A number of honeycomb lattice structures have been identified experimentally, including the spin-$\frac{1}{2}$ compound InCu$_{\frac{2}{3}}$V$_{\frac{1}{3}}$O$_3$,~\cite{Moller_2008, Yehia_2010} the spin-$\frac{3}{2}$ systems Bi$_3$Mn$_4$O$_{12}$(NO$_3$)~\cite{Smirnova_2009, Matsuda_2010} and $\beta-$CaCr$_2$O$_4$~\cite{Damay_2010, Damay_2011} as well as other systems with larger spins like Ba$Ln_2$O$_4$,~\cite{Doi_2006} Eu$Ln_2$O$_4$~\cite{Ofer_2011} and \slo.~\cite{Karunadasa_2005}

Bulk property measurements of geometrically frustrated systems usually reveal anomalous behaviour, such as the presence of phase transitions at temperatures much lower than expected from the strength of the exchange interactions. 
Magnetic susceptibility, $\chi(T)$, measured on powder samples of all the \slo\ compounds has revealed a disparity in the measured Curie-Weiss constants, $\theta_{CW}$, and the lack of long-range order down to 1.8~K.~\cite{Karunadasa_2005} 
\seo\ and \sho\ show cusps in single crystal $\chi(T)$ measurements at $T$~=~0.62~K for \sho~\cite{Hayes_2012} and at $T_N$~=~0.75~K for \seo,~\cite{Petrenko_2008, Hayes_2012} and a $\lambda$-anomaly is observed in the specific heat of \syo\ at $T_N$~=~0.92~K.~\cite{Quintero_2012}
In contrast, \sdo\ does not show any anomalies in $\chi(T)$ down to 0.5~K,~\cite{Hayes_2012} and no $\lambda$-type transitions have been observed in heat capacity measurements down to 0.39~K in zero applied field.~\cite{Cheffings_2013}
Neutron diffraction on powder samples of \sdo\ reveals only broad diffuse scattering peaks and thus further corroborates the apparent lack of long-range order at least down to 20~mK for this compound.~\cite{Petrenko_2007}

Low-temperature neutron diffraction studies of \syo\ point to unusual magnetic behaviour, where the two inequivalent sites of the Yb$^{3+}$ ion do not appear to have the same value of the ordered moment.~\cite{Quintero_2012}
The magnetism in \seo\ is equally unconventional since powder neutron diffraction measurements indicate that the transition to N\'{e}el order involves only half of the Er$^{3+}$ sites (either the red or blue sites in the notation of Fig.~\ref{Lnsublattice}).~\cite{Petrenko_2008}
Polarized neutron studies on single crystals of \seo\ suggest that the long-range ordered ({\bf k}~=~0) magnetic structure coexists with a shorter-range incommensurate ordering that has a pronounced low-dimensional character at temperatures below $T_N$~=~0.75~K.~\cite{Hayes_2011}
Neutron diffraction on powder samples of \sho~\cite{Young_2012} appeared to indicate that a similar situation arises in this compound, with distinct long-range commensurate and shorter-range incommensurate magnetic structures present at low-temperatures.
As a result of refining the data collected at 0.045~K, a magnetic structure with the propagation vector {\bf k}~=~0 was proposed, with only one of the crystallographically inequivalent Ho$^{3+}$ ions carrying a significant magnetic moment.
All of the moments in the refined long-range structure point along the $c$ axis in ferromagnetic chains, with neighboring chains coupled antiferromagnetically.

In this paper we present the results of an investigation into the nature of the magnetism in single crystal samples of \sho\ at low-temperatures. 
We have measured neutron diffraction patterns for three orthogonal directions in the reciprocal space of \sho.
Our results indicate that in \sho\ \emph{two} distinct types of \emph{short-range} magnetic order arise from the two sites for the Ho$^{3+}$ ions.
In the ($hk0$) plane, the first type of diffuse magnetic scattering appears around positions with the propagation vector {\bf k}~=~0 below 0.7~K (with no long-range order present in contrast to previous measurements on powder samples of \sho). 
Whereas the second type of diffuse scattering appears as planes of scattering intensity at ($hk\pm\!\!\frac{l}{2}$). 
These planes are seen as ``rods'' of scattering intensity in both the ($h0l$) and ($0kl$) planes in reciprocal space at $Q$~=~(00$\frac{1}{2}$) and symmetry related positions.
Therefore the second type of short-range order present in this material is one-dimensional in real space. 
Thus two members of the \slo\ series, \sho\ and \seo, both show the coexistance of two types of magnetic ordering at low-temperatures, but \seo\ has long- and short-range components, whereas in \sho\ both types of magnetic correlations are short-ranged.

Careful studies of the correlation lengths ($L$) in \sho\ indicate that the diffuse scattering in the ($hk0$) plane is correlated to $L$~$\approx$~150~\AA\ in the \emph{a-b} plane at 0.18~K, and to $L$~$\approx$~190~\AA\ along the $c$ axis at 0.16~K. The scattering is mostly isotropic in reciprocal space.
The diffuse intensity planes have a correlation length of $L$~$\approx$~230~\AA\ along the $c$ axis at 0.15~K.
Finally by utilising an energy analyser we were able to deduce that the diffuse scattering in the ($hk0$) plane is purely elastic, while the diffuse scattering that appears in planes along ($00l$) is mostly elastic.

\section{Experimental Details}

Large volume single crystals of \sho\ were grown using the floating zone method, with details of the procedure reported elsewhere.~\cite{Balakrishnan_2009} 
The samples were aligned using the backscattering Laue technique and cut perpendicular to the principal crystal axes, within an estimated accuracy of $2^{\circ}$ for the D7 measurements and within $1^{\circ}$ for the WISH experiment. 
For the WISH experiment a single 0.9~g sample of \sho\ was fixed to an oxygen free copper sample holder using Kwikfill resin.
Two more 0.9~g samples of \sho\ (one for each orientation) were fixed to oxygen free copper sample holders using Kwikfill resin during the D7 experiment; with equivalent sample mountings used for background measurements. 
For the D10 experiment single crystal samples of \sho\ (weighing 0.3~g and 0.1~g) were fixed to oxygen free copper pins using Stycast resin.

Neutron diffraction measurements in the ($hk0$) scattering plane were made using the cold neutron WISH~\cite{Chapon_WISH} instrument at the ISIS facility (Rutherford Appleton Laboratory, UK) in a range of temperatures from 0.05 to 1.5~K. 
WISH is equipped with a continuous array of position sensitive $^3$He detectors that cover an angular range of 10$^{\circ} < 2\theta < 170^{\circ}$. 
This provides the substantial $Q$-space coverage required for single crystal experiments. 
Depending on the experimental requirements the WISH instrument resolution and flux can be tuned, and out of plane neutron detection is also possible.

The diffuse scattering spectrometer D7~\cite{Sharpf_1990, Sharpf_1993, Stewart_D7} (at the Institut Laue-Langevin in Grenoble, France) with {\it XYZ} polarization analysis was used to collect neutron scattering data for both ($h0l$) and ($0kl$) scattering planes in a range of temperatures from 0.055 to 4.5~K.  
The D7 instrument is equipped with 132 $^3$He detectors which cover an angular range of 4$^{\circ} < 2\theta < 145^{\circ}$. 
Throughout the experiment cold neutrons monochromated to a wavelength of 3.1~\AA\ were used, resulting in $Q$-space coverage of 0.14 to 3.91~\AA$^{-1}$.
A quartz standard was used for normalising the polarization efficiency and a vanadium standard was used for normalising the detector efficiency of the instrument. 
Scans were performed in which the crystal was rotated about the vertical axis in steps of 1$^{\circ}$.
Six measurements required to carry out full {\it XYZ} polarization analysis were performed at each temperature for both sample orientations.
The reciprocal space maps were constructed in the usual manner using standard D7 data reduction functions.

The high-flux single-crystal diffractometer D10 (located on a thermal neutron guide at the Institute Laue-Langevin in Grenoble, France) was used in the four-circle dilution refrigerator mode to collect precise neutron scattering data.
The sample pins were attached to the Eulerian cradle inside a helium flow cryostat with the dilution option to allow scanning along all reciprocal lattice directions at low-temperature.
A vertically focusing pyrolytic graphite (PG) (002) monochromator gave an incident wavelength of 2.36~\AA, with the half-wavelength contamination suppressed by a PG filter. 
A vertically focussing PG analyser was used to give improved resolution and suppressed background for some of the measurements. 
The two-dimensional area detector was used for all the measurements except for those made with the energy analyser, where a single $^3$He detector was used instead.
Measurements were made in the temperature range 0.15 to 10~K.
The integrated intensity of the magnetic diffuse scattering around the nuclear Bragg peaks was extracted in the usual manner.~\cite{Wilkinson_D10} 
The plane like diffuse scattering features, however, often had a broad intensity profile which covered a large proportion of the two-dimensional D10 detector. 
The selection of the integration box for these features required a compromise between getting adequate resolution and the correct representation of the diffuse scattering profile.

\section{Results and Discussion}
\subsection{(\emph{hk}0) plane}

\begin{figure}[tb]
   \centering
	\includegraphics[width=0.99\columnwidth]{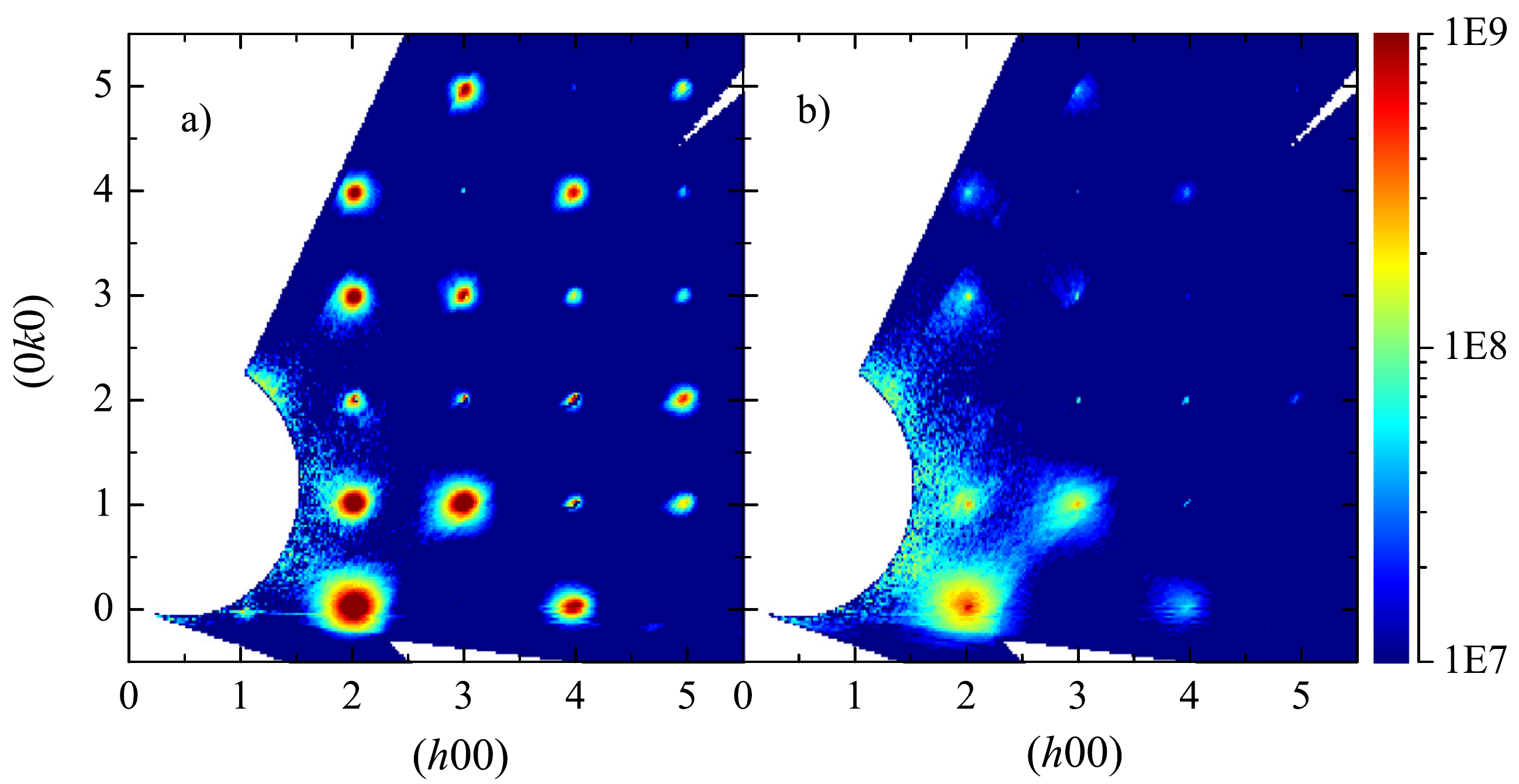}
	\caption{(Color online) The magnetic contribution to the scattering from \sho\ in the ($hk0$) plane at a) 0.055~K and b) 0.7~K. 
The magnetic component was isolated by subtracting a 1~K background from the single crystal neutron diffraction data collected using WISH, ISIS.~\cite{Chapon_WISH}}
   \label{hk0}
\end{figure}

\begin{figure}[tb]
  \centering
	\includegraphics[trim=1cm 0.5cm 2.3cm 2cm, clip=true, width=0.89\columnwidth]{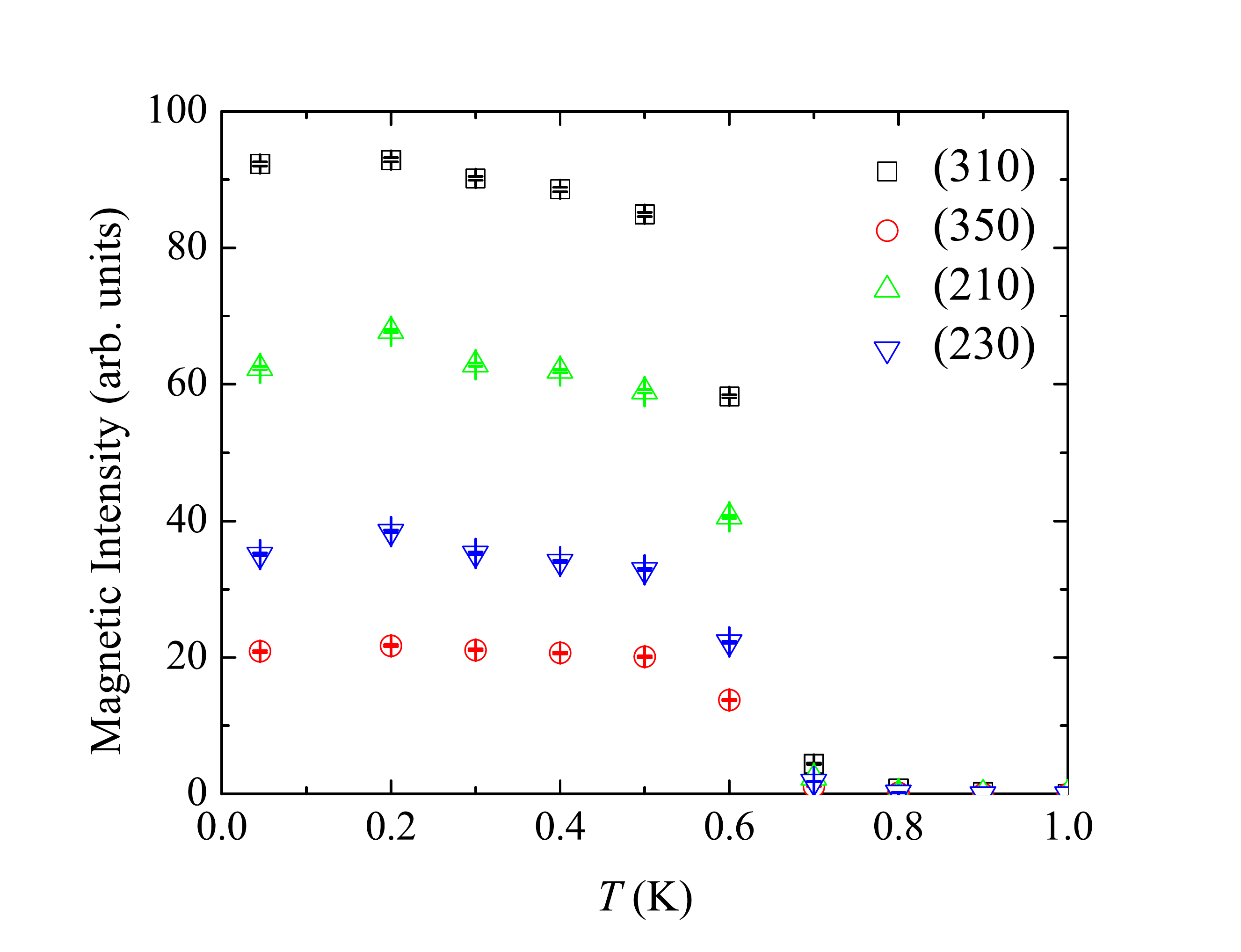}
	\caption{(Color online) Temperature dependence of several diffuse peaks intensities in the ($hk0$) plane as a function of temperature. The magnetic intensity was isolated by subtracting a 1~K background from data collected using WISH, ISIS.}
  \label{hk0_tdep}
\end{figure}

\begin{figure}[tb]
   \centering
	\includegraphics[trim=0.75cm 0.75cm 2.3cm 2cm, clip=true, width=0.99\columnwidth]{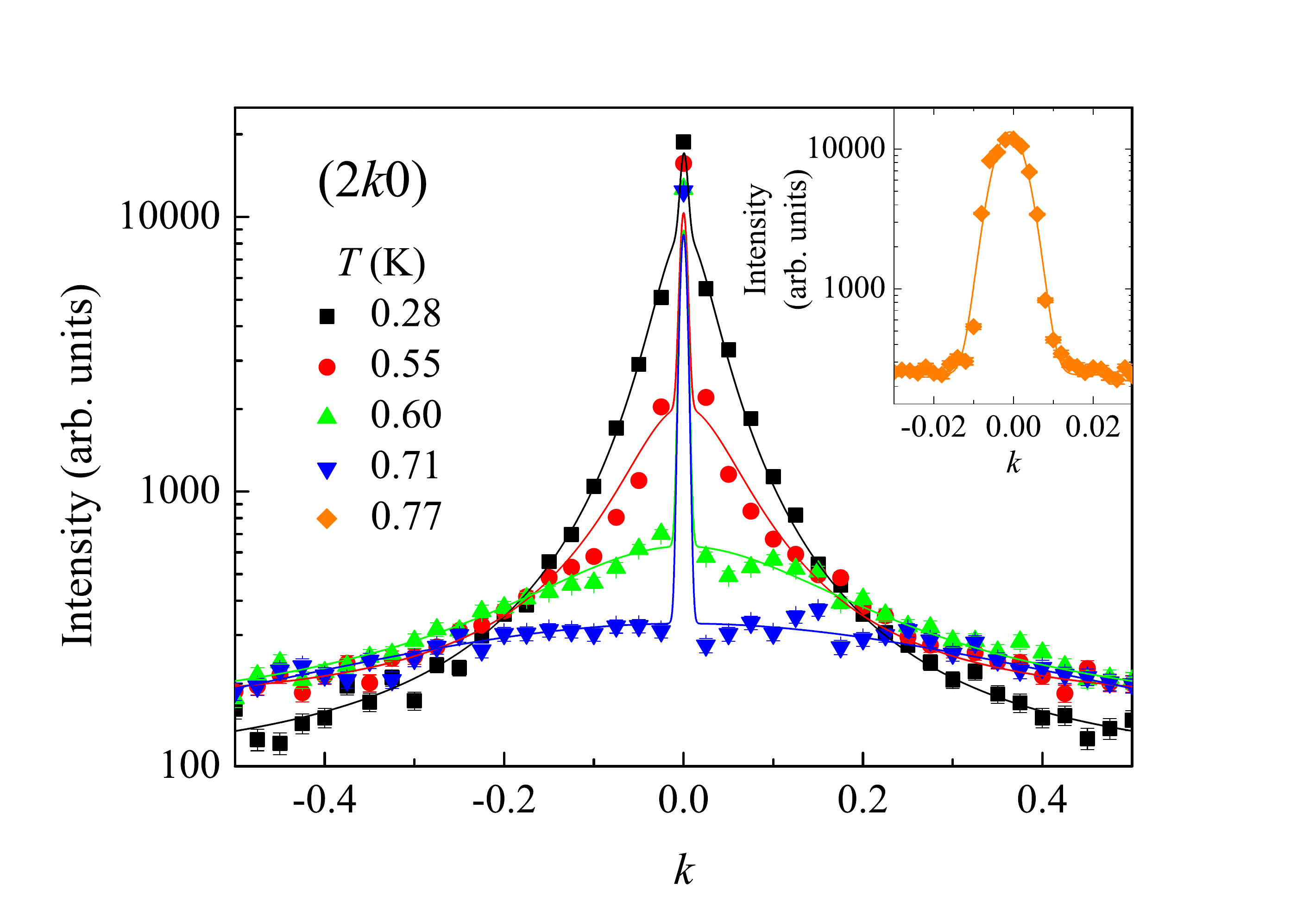}
	\caption{(Color online) Temperature dependence of the scattering intensity of the (200) reflection from \sho.
	The data are fitted using Gaussian (nuclear) and Lorentzian (magnetic) components, and was collected on the D10 instrument at the ILL.
	The inset shows the data collected for the nuclear (200) peak at high temperature with high resolution and the fitted Gaussian distribution.}
   \label{200peak}
\end{figure}

\begin{figure}[tb]
   \centering
	\includegraphics[trim=0.75cm 0.75cm 2.3cm 2cm, clip=true, width=0.99\columnwidth]{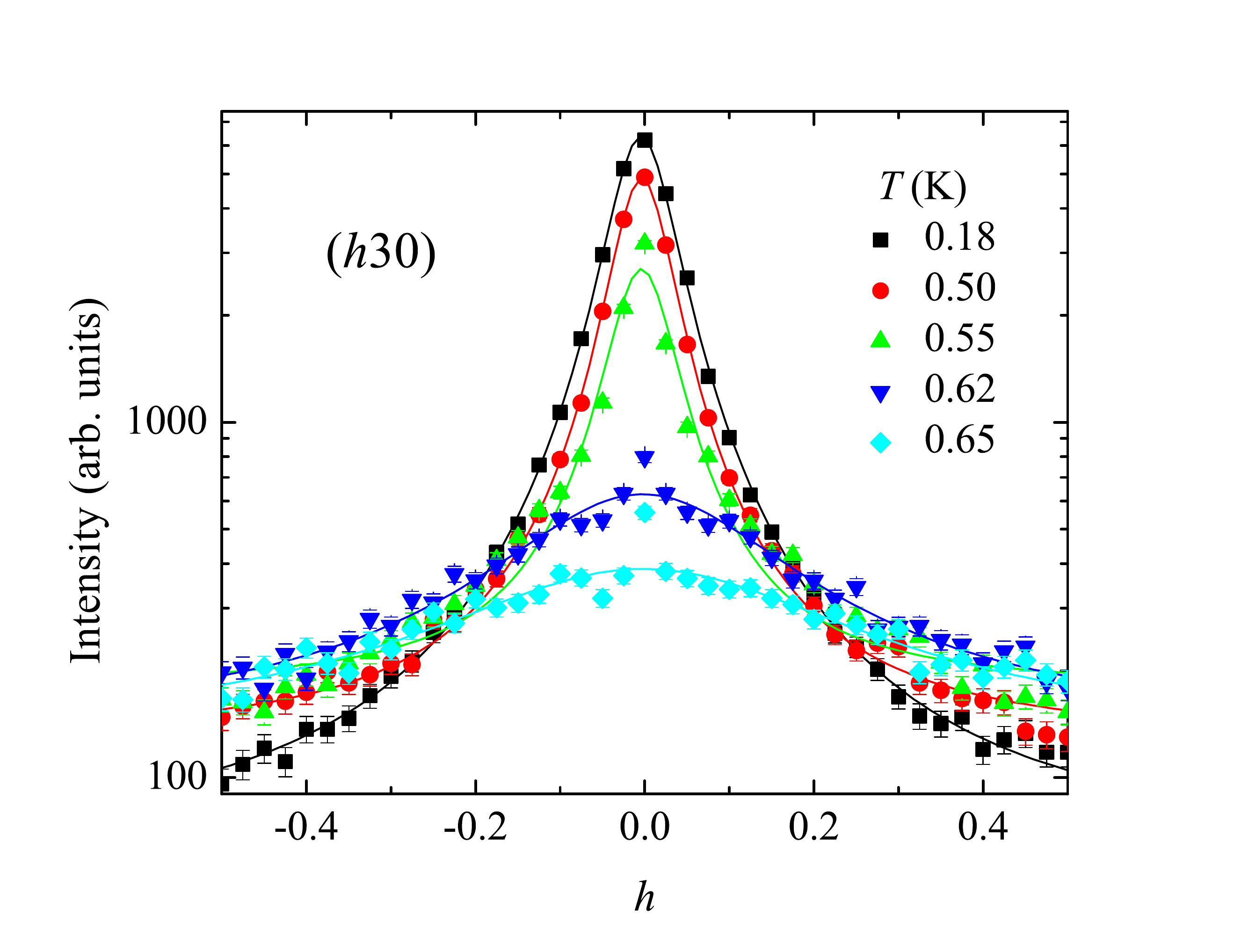}
	\caption{(Color online) Temperature dependence of the scattering intensity of the (030) reflection from \sho.
	The data are fitted using the Lorentzian distribution and was collected using the D10 instrument at the ILL.}
   \label{030peak}
\end{figure}

\begin{figure}[tb]
   \centering
	\includegraphics[trim=0.75cm 0.75cm 2.3cm 2cm, clip=true, width=0.99\columnwidth]{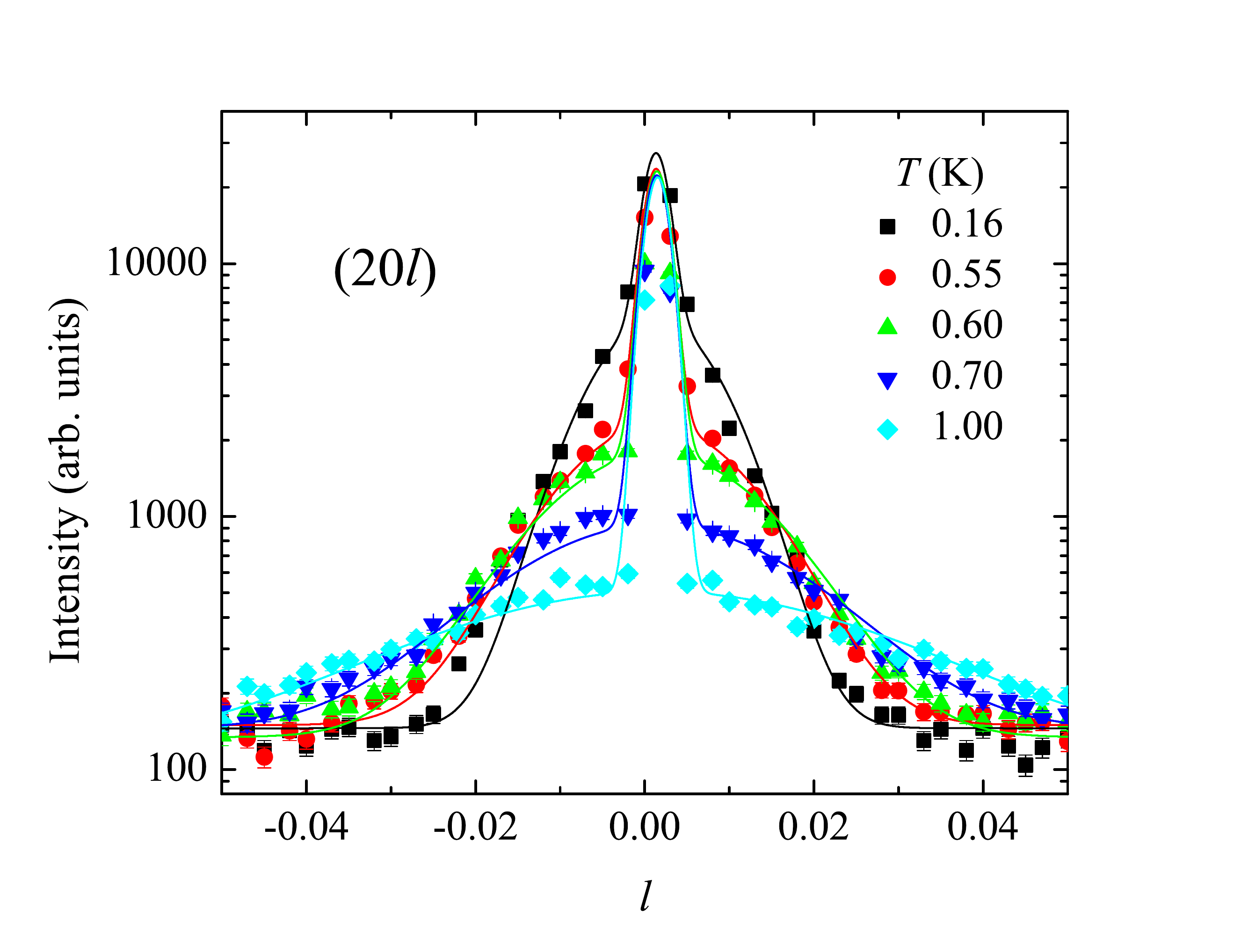}
	\caption{(Color online) Temperature dependence of the scattering intensity of the (200) reflection from \sho.
	The data are fitted using two Gaussian components (one for nuclear and one for magnetic contributions) at all temperatures.
	The data were collected using the D10 instrument at the ILL.
}
   \label{20lpeak}
\end{figure}

From earlier measurements on a powder sample of \sho, {\bf k}~=~0 magnetic Bragg peaks were expected in the ($hk0$) scattering plane,~\cite{Young_2012} similar to what has previously been observed for \seo.~\cite{Hayes_2011}
However, after cooling down the single crystal sample to 0.055~K, no long-range order was found in the WISH measurements. 
Fig.~\ref{hk0} shows the observed magnetic scattering (which was isolated by subtracting a 1~K background from the low-temperature data). 
From the single crystal data it is apparent that in the ($hk0$) plane diffuse magnetic scattering appears around the {\bf k}~=~0 positions and that its intensity increases rapidly below 0.7~K.
Waiting at low-temperature for a long time for the sample to thermalise made no measurable difference to the intensity or width of the diffraction patterns. 

It is important to note that in \sho\ at temperatures below 0.7~K the diffuse scattering forms the same \emph{lozenge} pattern that is seen at low-temperatures in \seo.~\cite{Hayes_2011} 
This is a result of very similar positions of the magnetic $Ln^{3+}$ ions in the respective unit cells for both compounds.

\begin{figure}[tb]
   \centering
	\includegraphics[trim=0.25cm 0.5cm 0.25cm 1.5cm, clip=true, width=0.99\columnwidth]{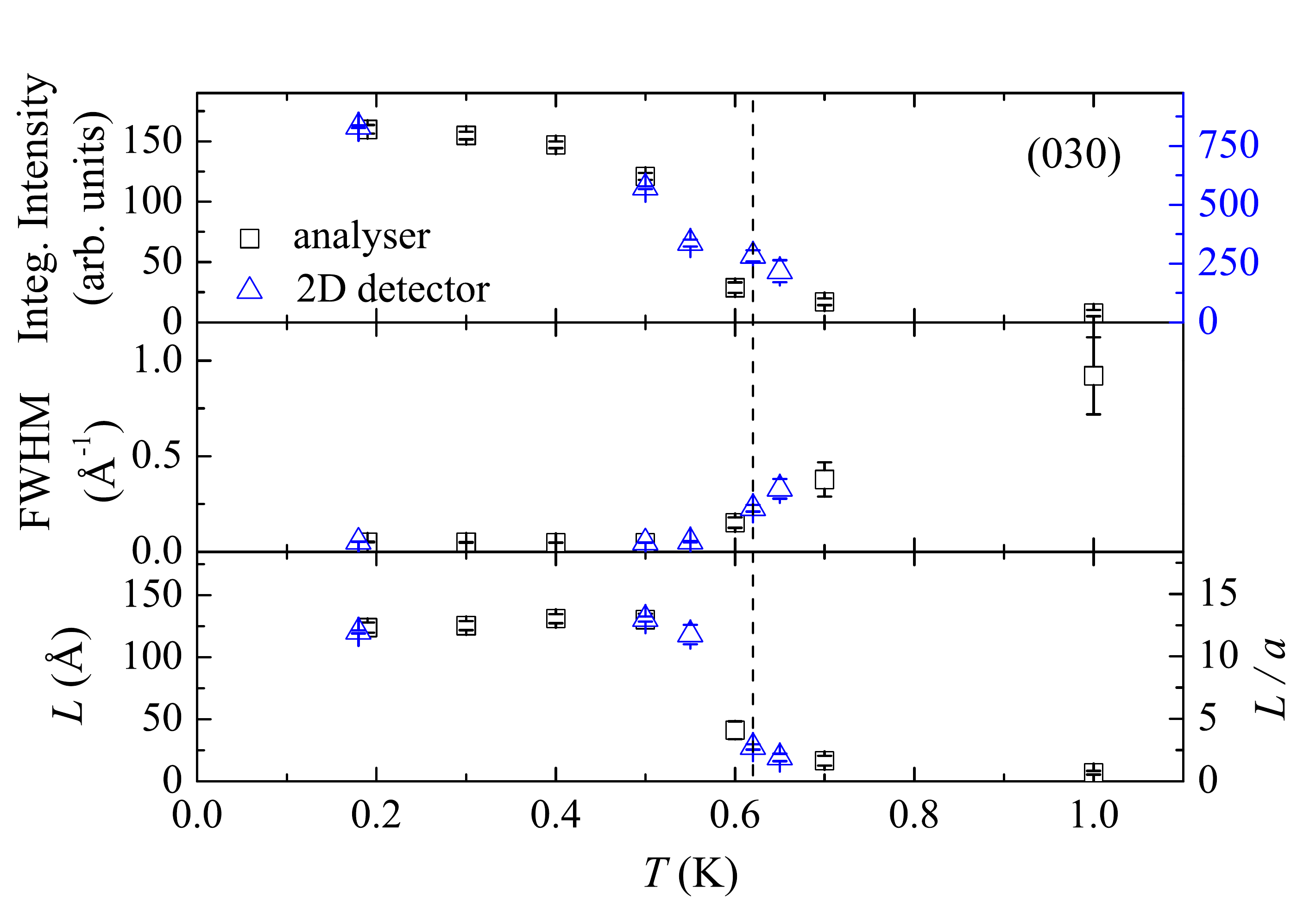}
	\caption{Temperature dependence of the integrated intensity and the FWHM along $h$ of the (030) reflection from \sho, as well as the calculated correlation length, $L$, along the $a$ axis. 
The integrated intensity and the FWHM were extracted by fitting a Lorentzian distribution to data collected using the D10 instrument at the ILL in both the area detector and the energy analyser configurations.
For the top panel, the righthand axis shows the intensity scale for the area detector instrument configuration.
The righthand axis of the bottom panel is a conversion to how many unit cells the magnetic order is correlated to.
The dashed line indicates 0.62~K - the temperature at which a cusp is observed in single crystal magnetic susceptibility measurements.~\cite{Hayes_2012}
}
   \label{hk0_2ndorder}
\end{figure}

The positions of the observed broad diffuse scattering peaks in the ($hk0$) scattering plane are in agreement with the positions previously reported for a {\bf k}~=~0 structure obtained from the powder refinement. 
However, from the single crystal experiment no long-range order is actually observed even though the onset of the diffuse scattering in the ($hk0$) plane appears to go through a second-order phase transition (see Fig.~\ref{hk0_tdep} and Fig.~\ref{hk0_2ndorder}), and can also be observed as a clear cusp in single crystal magnetic susceptibility measurements.~\cite{Hayes_2012}

Although the $Q$ resolution of WISH is good the sample was kept stationary during the experiments and in order to measure correlation length more accurately, high resolution scans across the nuclear and magnetic reflections in the ($hk0$) plane were performed using the D10 instrument.
Temperature dependence of the scattering intensity for two different types of reflections in the ($hk0$) plane of \sho\ are shown in Fig.~\ref{200peak}, Fig.~\ref{030peak} and Fig.~\ref{20lpeak}.
The data in Fig.~\ref{200peak} have been fitted with Gaussian and Lorentzian components, to account for the nuclear (temperature independent) and magnetic scattering.~\cite{2Gauss}
The nuclear (Gaussian) component was fitted to data collected for (200) at higher temperature with high resolution, and this is shown in the inset to Fig.~\ref{200peak}.
The data in Fig.~\ref{030peak} have been fitted using the Lorentzian distribution only, since this is a forbidden reflection in \emph{Pnam} symmetry, and purely magnetic scattering is expected.
The data in Fig.~\ref{20lpeak} have been fitted using two Gaussian components (one for nuclear and one for magnetic contributions) at all temperatures.~\cite{2Gauss}

From the FWHM of the fitted Lorentzian curves to the diffuse scattering, the correlation lengths were estimated using $L$~(\AA)~=~$2\pi$(FWHM (\AA$^{-1}))^{-1}$.~\cite{CorrCalc}
At the lowest temperature, the magnetic order is correlated to $L$~$\approx$~150~\AA\ in the \emph{a-b} plane, and to $L$~$\approx$~190~\AA\ along the $c$ axis.
The integrated intensity and FWHM along $h$, and the calculated correlation length along the $a$ axis of the (030) reflection is plotted as a function of temperature in Fig.~\ref{hk0_2ndorder}.
The data for the (030) reflection in Fig.~\ref{hk0_2ndorder} were collected using both the two-dimensional detector and energy analyser D10 instrument configurations.
A detailed comparison of these two types of measurements is presented later, see section~\ref{analyser}.
Overall, the diffuse scattering in the ($hk0$) plane shows second-order phase transition type behaviour, with a marked increase in correlation length below 0.62~K - the temperature at which a cusp was observed in single crystal magnetic susceptibility measurements.~\cite{Hayes_2012}

Finally, it is important to note that with the improved instrumental resolution offered by D10, it is possible to see that there is a small amount of scattering intensity at positions forbidden by the $Pnam$ symmetry.
This suggests that the proposed space group is not entirely appropriate for the \slo\ family of compounds, although the relative intensity of the ``forbidden" peaks is systematically low compared to the intensity of the allowed nuclear peaks reflections, therefore much more accurate collection of the intensity of all the peaks is required in order to solve the crystal structure.

\subsection{(\emph{h}0\emph{l}) and (0\emph{kl}) planes}

\begin{figure*}[tb]
	\centering
		\includegraphics[width=1.99\columnwidth]{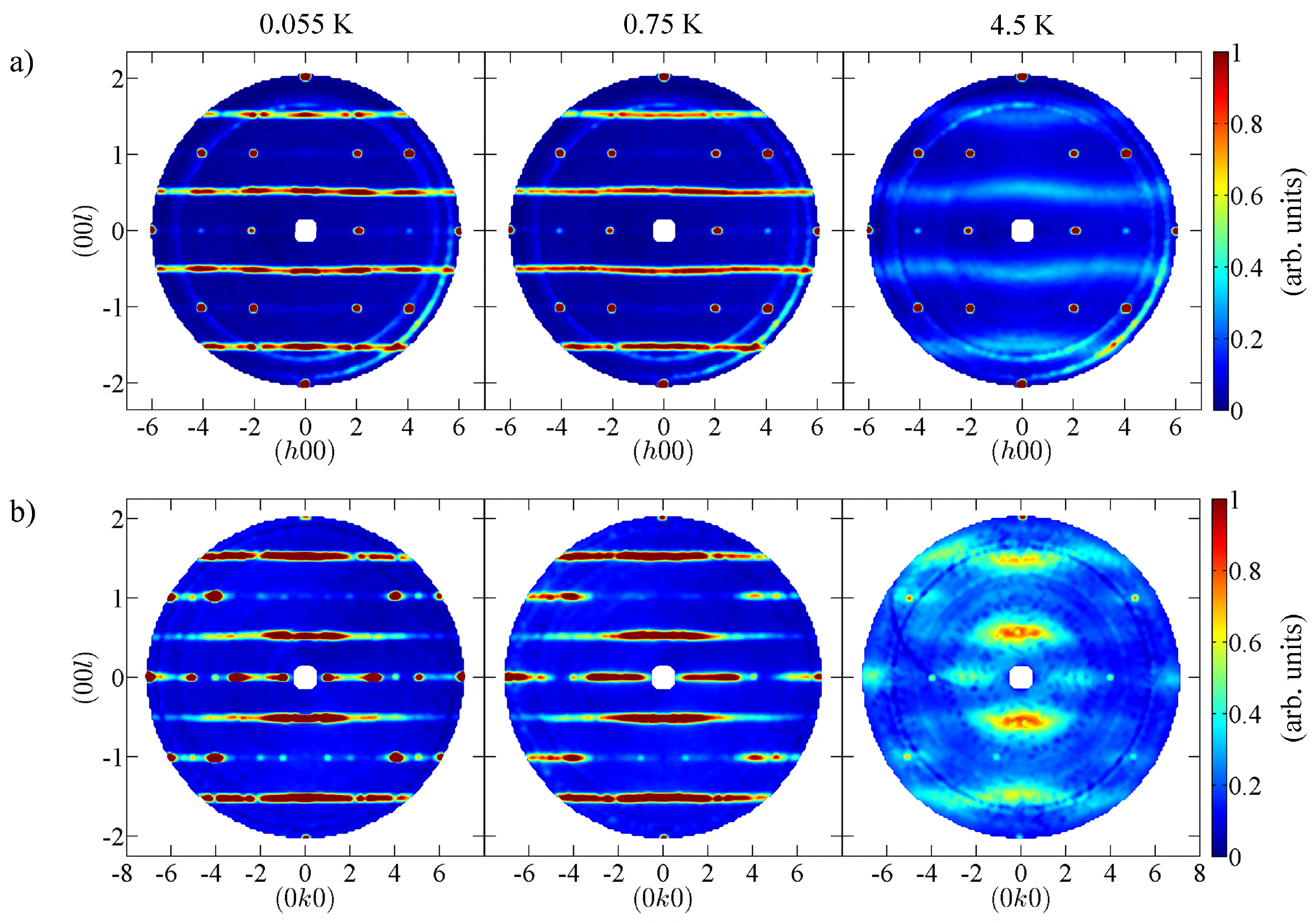}
		\caption{(Color online) Intensity maps of the scattering from SrHo$_2$O$_4$ at different temperatures, made from data collected using the D7 instrument \cite{Sharpf_1990, Sharpf_1993, Stewart_D7} at the ILL. 
(a)~The $Z$~non-spin-flip scattering in the ($h0l$) plane and (b)~the $Z$~spin-flip scattering intensity in the ($0kl$) plane. 
For both sample orientations ``rods'' of scattering intensity at ($h0\frac{1}{2}$) and ($0k\frac{1}{2}$) and symmetry related positions are visible at low temperature.
In the ($0kl$) plane at 0.055~K diffuse magnetic peaks associated with the {\bf k}~=~0 structure are also visible.
}
   \label{D7-2planes}
\end{figure*}

The intensity maps shown in Fig.~\ref{D7-2planes} summarise the main results of the single crystal neutron diffraction experiments to investigate the second type of diffuse scattering present in \sho.
The polarized neutron experiments were configured to measure two independent components of the scattering function, here called the Non-Spin-Flip (NSF) and Spin-Flip (SF) channels.
The $Z$ polarization direction in our measurements coincides with the vertical axis, and hence the sum of $Z$~NSF and $Z$~SF measurements is what would be observed in unpolarized neutron experiments.
The standard data reduction formulas utilising all of the \emph{XYZ} polarized measurements to calculate the purely magnetic contribution to the scattering \cite{Sharpf_1993, Stewart_D7} cannot be applied to the \sho\ system, because it is not an isotropically magnetic crystal at low-temperatures.
Instead, Fig.~\ref{D7-2planes}(a) shows the intensity measured in the ($h0l$) plane $Z$~NSF measurement (here, the $Z$ polarization direction coinsides with the $b$ axis).
No extra magnetic features associated with the ($hk\pm\!\!\frac{l}{2}$) planes are observed in the $Z$~SF cross-section for this sample orientation.
Fig.~\ref{D7-2planes}(b) shows the data for the ($0kl$) plane $Z$~SF measurements (here, the $Z$ polarization direction coinsides with the $a$ axis).
Similarly, no extra magnetic features are observed in the $Z$~NSF cross-section for the second sample orientation.
Nonetheless, all of the measured $Z$ polarization cross-sections are given in the supplimentary material.
This method of looking at the magnetic intensity ignores the contribution of components of the moments orthogonal to the $Z$ axis in the ($h0l$) plane and parallel to the $Z$ axis for the ($0kl$) plane, however our complimentary measurements with unpolarized neutrons on the E2 instrument~\cite{Young_E2} also show no new information on the magnetic structure compared to the D7 results.

Measurements of the ($h0l$) and ($0kl$) scattering planes reveal broad ``rod''-like features at half-integer positions along the $l$-axis observed at all measured temperatures.
The position of the rods along $l$ suggests that the unit cell for this magnetic structure is doubled along the $c$ axis. 
In the ($h0l$) plane, the rods extend along the $h$-direction which implies that the magnetic moments are only weakly correlated along the $a$ axis. 
Also, in the ($0kl$) plane, the rods extend along the $k$-direction, so the magnetic moments are only weakly correlated along the $b$ axis.
The presence of the ``rods'' along two orthogonal directions means that the diffuse scattering is a two-dimensional structure in reciprocal space.
Hence the correlations of this type of short-range order in \sho\ are $one$-dimensional in real space.

At the lowest temperature of 0.055~K the ``rods'' of diffuse scattering are most intense and approximate straight lines perpendicular to the $l$-direction.
In the ($h0l$) plane areas of greater intensity along the rod can also be seen at even integer $h$.
There is no pronounced decay of intensity for the ``rods'' with increasing $h$, $k$ or $l$ than what is likely due to the magnetic form factor.
On warming the system to 0.75~K the one-dimensional scattering in both the ($h0l$) and ($0kl$) planes is still very intense, and for the ($0kl$) plane it is little different to what is observed at base temperature.
However, in the ($h0l$) plane there is no longer any modulation of the intensity along the rods.
Thus the onset of the diffuse scattering in the ($hk0$) plane below 0.7~K seems to only affect the modulation of the intensity along the rods in the ($h0l$) plane, with no other effect on the diffraction pattern of the second diffuse phase. 
In order to explain this lack of coupling between the two types of short-range ordering, we suggest that each of the crystallographically inequivalent Ho$^{3+}$ sites in \sho\ contributes to only one of the two coexisting structures. 
At much higher temperatures, $T=4.5$~K, there is still some intensity in the diffuse component of the total scattering, for example shown in the righthand panels of Figure~\ref{D7-2planes}.
The ``rods'' are now much wider along the $l$-direction, and they appear to be a lot less straight.
The one-dimensional order persists over a wide temperature range, and the onset of these correlations may be observed with bulk heat capacity measurements (as the broad peak around 3.5~K shown in the inset of Fig.~3 in Ref.~\citenum{Ghosh_2011}).

From the $Z$ polarization direction NSF and SF data, some conclusions can be drawn about the nature of this diffuse magnetic scattering.
In the ($0kl$) measurement the diffuse scattering appears at the ($0k\frac{1}{2}$) and symmetry equivalent positions only in the SF channel. 
This demonstrates that the spin structure has no components parallel to the $Z$ direction which coincides with the crystallographic direction $a$ in this measurement, \emph{i.e.} all of the magnetic moments lie in the \emph{b-c} plane.
In the ($h0l$) measurement the diffuse scattering appears at the ($h0\frac{1}{2}$) and symmetry equivalent positions only in the NSF measurement, which indicates that in this low-dimensional magnetic structure all of the spins lie \emph{parallel} with the $b$ axis.

The diffuse scattering peaks associated with the {\bf k}~=~0 magnetic structure are only visible in the $Z$ SF data for both the ($h0l$) and ($0kl$) scattering planes.
This means that all of the spins that give rise to the {\bf k}~=~0 scattering are \emph{parallel} with the $c$ axis.
This is in agreement with the previously proposed magnetic structure that was established from a refinement performed using powder neutron diffraction data.~\cite{Young_2012}
Single ion anisotropy, due to the crystal field, may be responsible for the magnetic moments on the two crystallographically inequivalent sites pointing along two orthogonal principal axes.

\begin{figure}[tb]
   \centering
	\includegraphics[trim=1.25cm 0.75cm 2.3cm 2cm, clip=true, width=0.99\columnwidth]{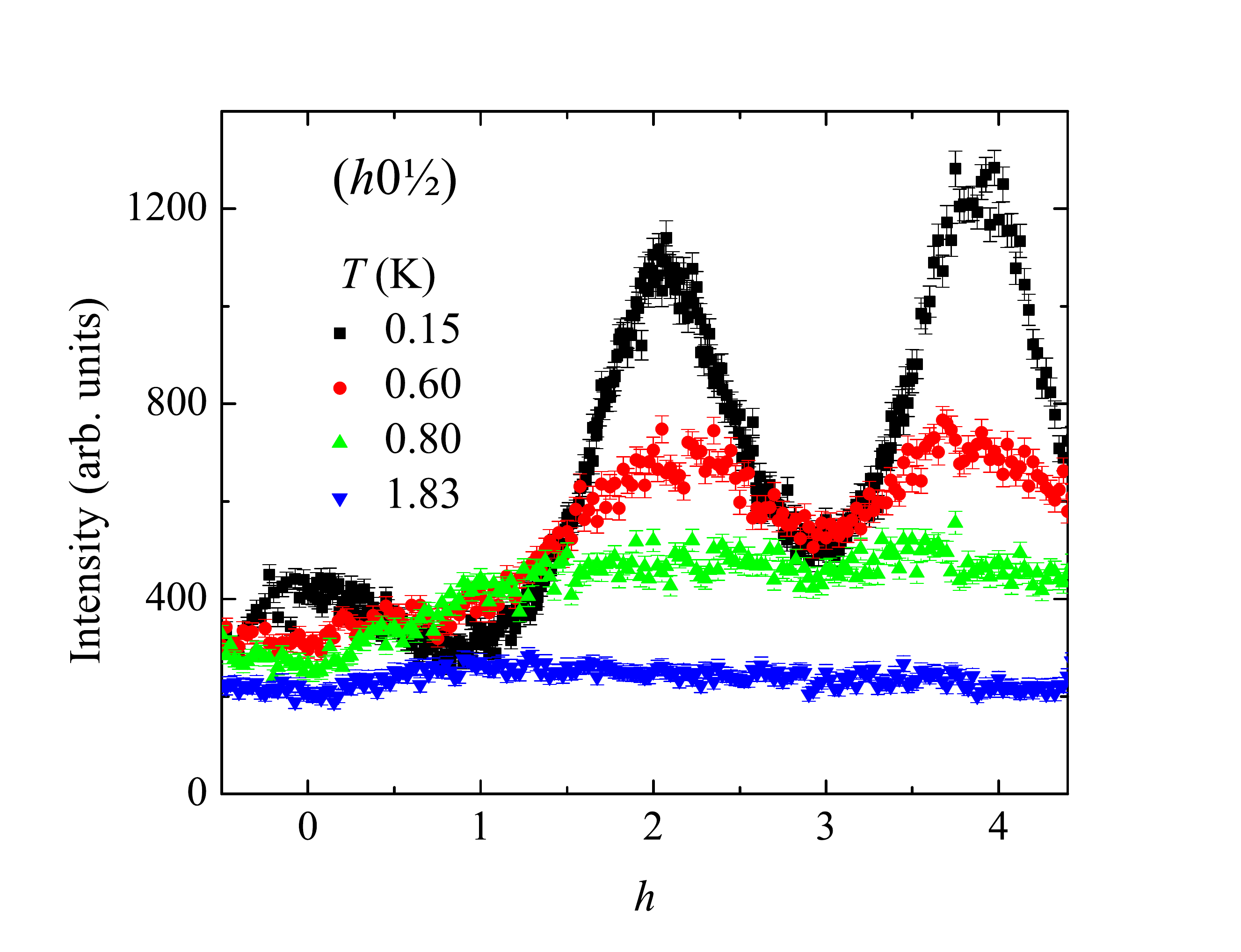}
	\caption{(Color online) $Q$-dependence of the scattering intensity along one of the ``rods'' in the ($h0l$) plane measured at different temperatures for the \sho\ crystal using the D10 instrument at the ILL.
	At the lowest temperature of 0.15~K, the scattering is most intense at even integer values of~$h$. 
	As the temperature is increased to 1.83~K, the diffuse scattering is still measurable above background, but the $Q$-dependence of the scattering intensity along the ``rod'' is no longer present.}
   \label{diffTdep1}
\end{figure}

\begin{figure}[tb]
   \centering
	\includegraphics[trim=1.25cm 0.75cm 2.3cm 2cm, clip=true, width=0.99\columnwidth]{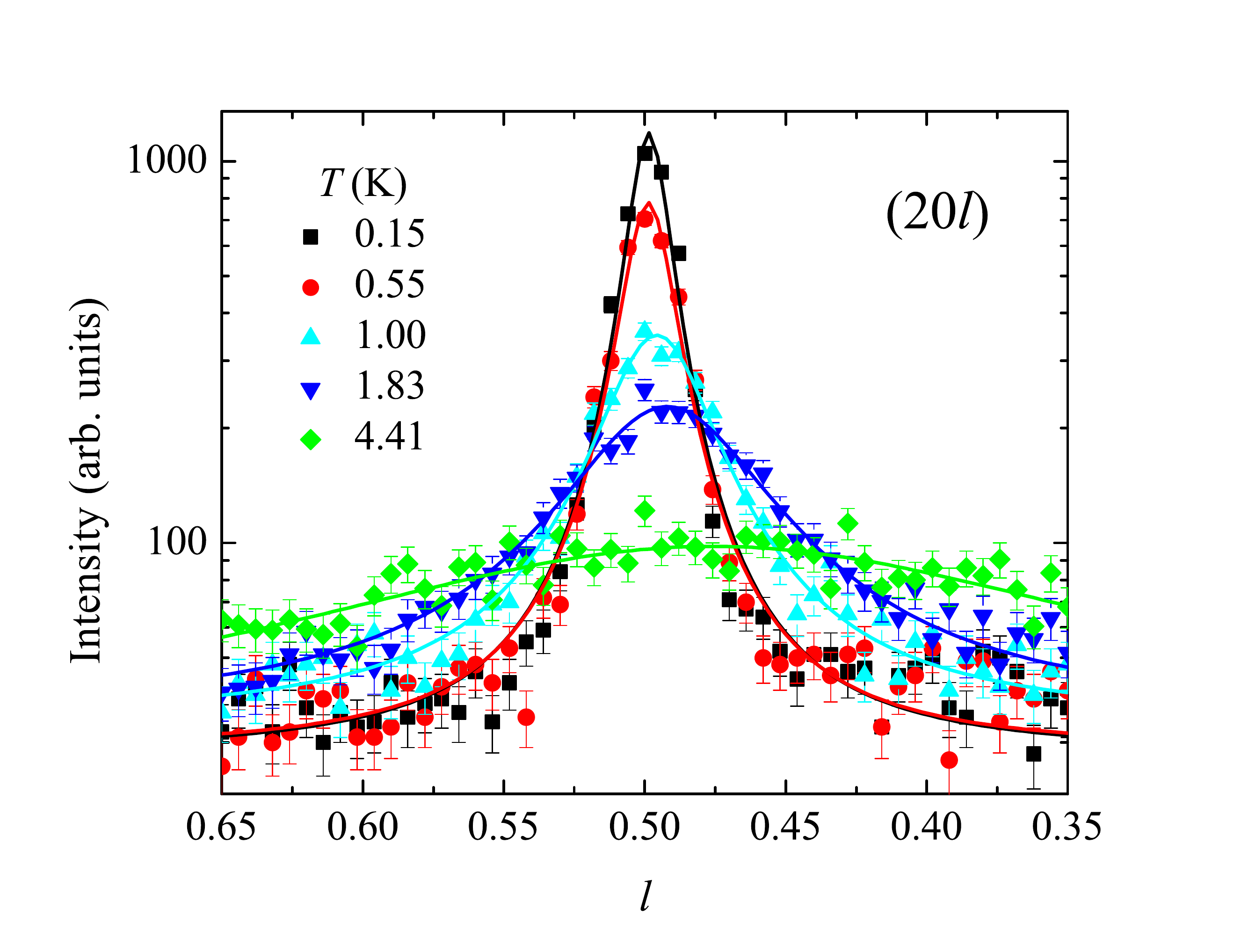}
	\caption{(Color online) Temperature dependence of the integrated intensity of the diffuse magnetic scattering around (20$\frac{1}{2}$), measured using the D10 instrument at the ILL.
	The data have been fitted using a Lorentzian distribution.}
   \label{diffTdep2}
\end{figure}

\begin{figure}[tb]
   \centering
	\includegraphics[trim=0.0cm 0.0cm 0.2cm 1.25cm, clip=true, width=0.99\columnwidth]{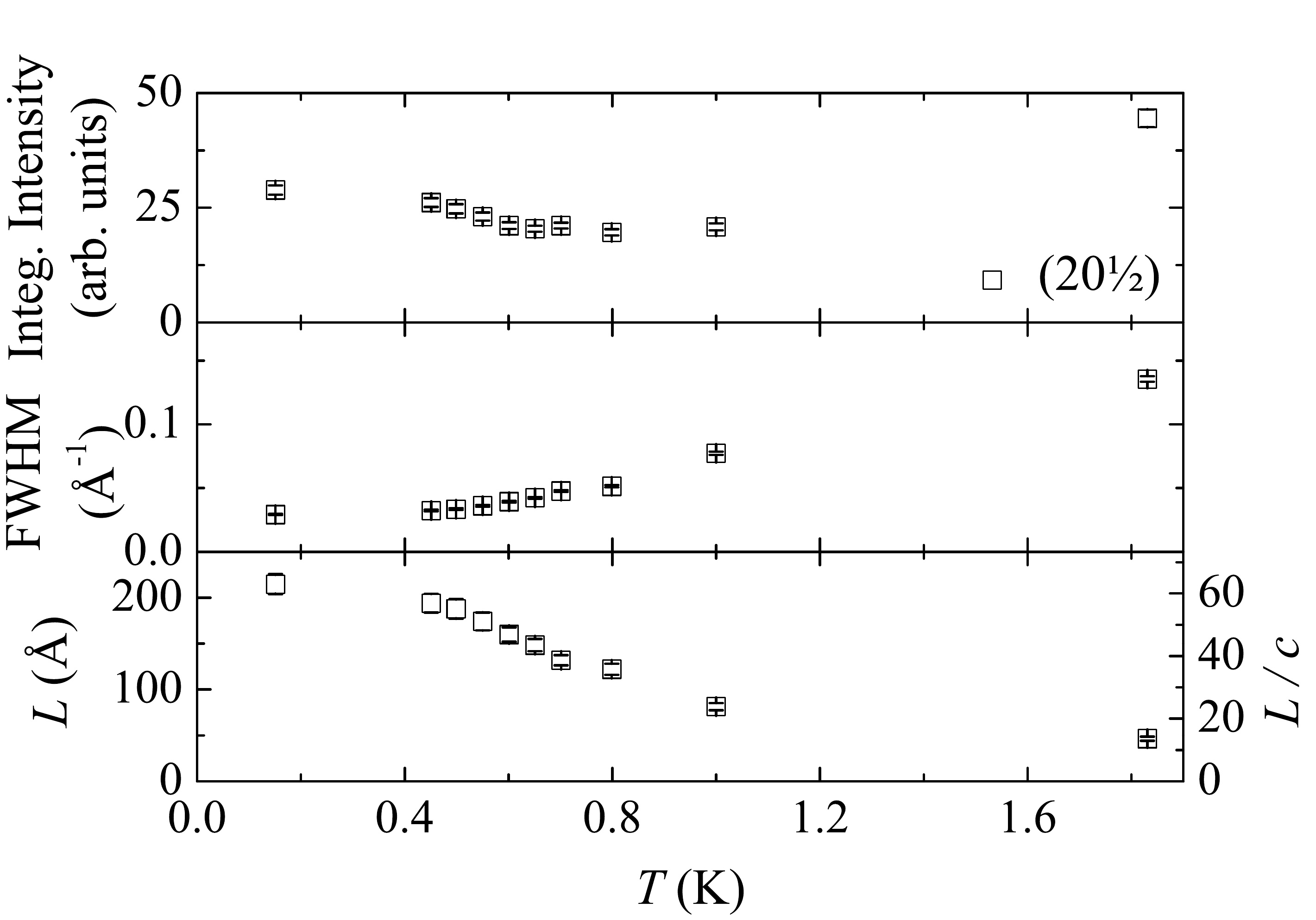}
	\caption{Temperature dependence of the integrated intensity and the FWHM along $l$ of the (20$\frac{1}{2}$) reflection from \sho. The calculated correlation length, $L$, along the $c$ axis is shown in the bottom panel, with the righthand axis giving a conversion to how many unit cells the magnetic order is correlated to.
The integrated intensity and the FWHM were extracted by fitting a Lorentzian distribution to data collected using the D10 instrument at the ILL.
}
   \label{20L}
\end{figure}

The $Q$ resolution of D7 does not allow for the precise investigation of the structure of the planes of diffuse scattering, so the D10 diffractometer was used to further examine the ``rod'' like features.
Scans along the rods at 0.15~K indicate that there is a pronounced $Q$-dependence to the scattering intensity, and to illustrate this, data for the ($h0\frac{1}{2}$) rod is shown in Fig.~\ref{diffTdep1}.
At the lowest temperature the scattering appears to be most intense near the integer values of $h$.
Upon raising the temperature the $Q$-dependence becomes less pronounced, but the scattering intensity remains measurable above background even at temperatures as high as 1.8~K, as expected from the D7 measurements.
In order to have a complete picture of the one-dimensional scattering, scans across the ``rods'' along the $l$-direction were also performed.
The temperature dependence of the integrated intensity of the diffuse magnetic scattering across ($20l$) is shown in Fig.~\ref{diffTdep2} as an example.
At the lowest temperature the moments are highly correlated around $l$~=~$\frac{1}{2}$.
Upon raising the temperature the scattering becomes less intense and less symmetrical about $l$~=~$\frac{1}{2}$.
Unlike the \emph{lozenge}-type diffuse scattering seen in the ($hk0$) plane, the diffuse scattering intensity seen in the ($h0l$) and ($0kl$) planes appears to be a smooth function of temperature, with no well-defined transition, just a gradual building-up of one-dimensional correlations.

From the FWHM of the fitted Lorentzian curves to the diffuse scattering, the correlation lengths were estimated as described previously.~\cite{CorrCalc2}
At the lowest temperature, the ($hk\pm\!\!\frac{l}{2}$) planes of diffuse scattering are correlated to $L$~$\approx$~230~\AA\ along the $c$ axis.
The integrated intensity and FWHM along $l$, and the calculated correlation length along the $c$ axis of the (20$\frac{1}{2}$) reflection are plotted as a function of temperature in Fig.~\ref{20L}.
The diffuse scattering does not show any abrupt transition, demonstrating a gradual build-up of correlations with decreasing temperature.

It is interesting to note that this type of diffuse scattering was expected from the previously collected diffraction data on powder samples of \sho.~\cite{Karunadasa_2005, Young_2012}
Broad features were seen at $Q$~=~(00$\frac{1}{2}$) and symmetry related positions.
However, from the powder data alone it could not be established whether the diffuse scattering was one- or two-dimensional.

\subsection{Energy analysis}
\label{analyser}

\begin{figure}[tb]
   \centering
	\includegraphics[trim=0.25cm 0.25cm 1.5cm 1.5cm, clip=true, width=0.99\columnwidth]{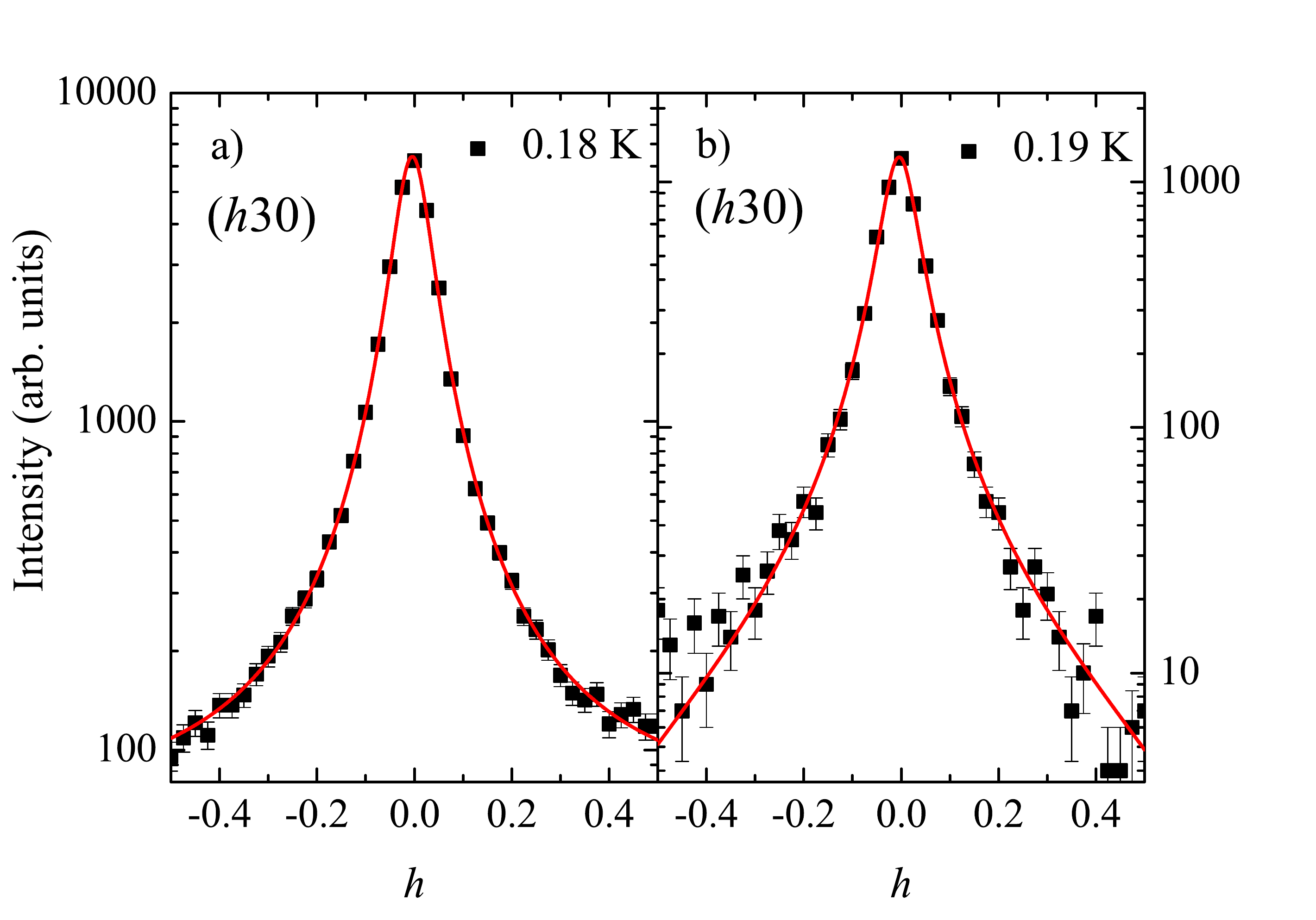}
	\caption{(Color online) Comparison of the magnetic contribution to the scattering from \sho\ when scanning across the (030) peak in a) 2-dimensional area detector configuration and b) energy analyser configuration. 
The peaks are fitted with a Lorentzian distribution, where the FWHM are a) (0.052~$\pm$~0.001)~\AA$^{-1}$ and b) (0.049~$\pm$~0.001)~\AA$^{-1}$.}
   \label{ANApeak}
\end{figure}

\begin{figure}[tb]
   \centering
	\includegraphics[trim=0.25cm 0.25cm 1.5cm 1.5cm, clip=true, width=0.99\columnwidth]{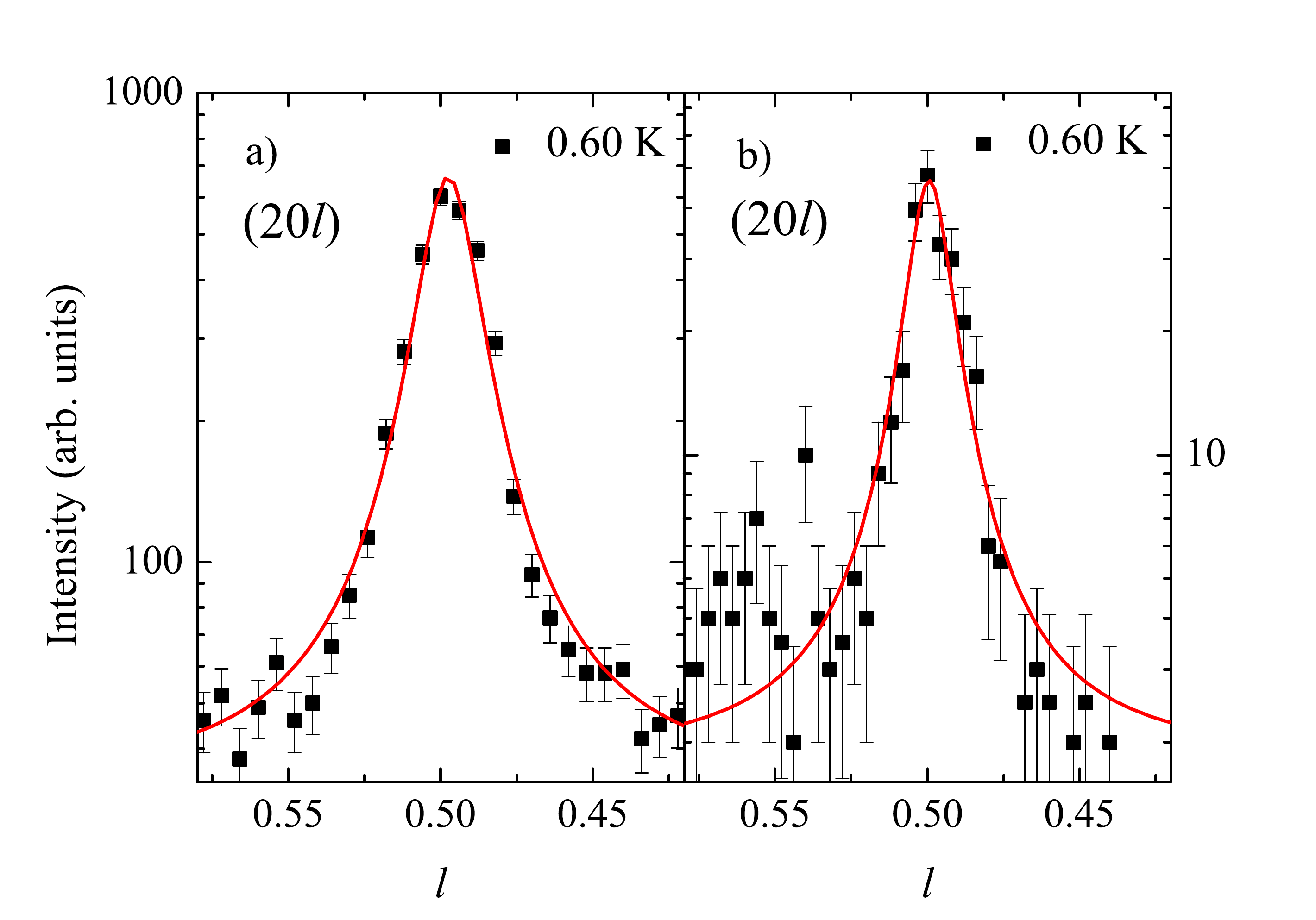}
	\caption{(Color online) Comparison of the diffuse magnetic scattering when scanning along (20$l$) in a) 2-dimensional area detector configuration and b) energy analyser configuration. 
When the data are fitted with a Lorentzian distribution, where the FWHM are a) (0.039~$\pm$~0.002)~\AA$^{-1}$ and b) (0.029~$\pm$~0.003)~\AA$^{-1}$.}
   \label{ANAdiff}
\end{figure}

Using the D10 instrument with the vertically focussing PG analyser both types of diffuse scattering found in \sho\ were investigated.
Comparison of the data collected with the analyser and with the 2-dimensional area detector data for the (030) magnetic reflection at the lowest temperature is shown in Fig.~\ref{ANApeak}.
The statistics are much lower for the data collected in the analyser configuration.
The peaks are fitted with a Lorentzian distribution, and the resulting FWHM for the two datasets is in reasonable agreement.
This indicates that the diffuse scattering that appears around the peaks in the ($hk0$) plane is elastic within the energy resolution of the analyser.

Scans along ($00l$) at several integer positions of $h$ were performed to check the diffuse scattering profile of the planes with an improved energy resolution.
The comparison of the data collected using the 2-dimensional area detector and that using the analyser configuration for the (20$\frac{1}{2}$) reflection at 0.60~K is shown in Fig.~\ref{ANAdiff}.
Here, the FWHM is fairly close for both datasets, taking into account that the statistics for data collected with the analyser configuration are significantly lower.
This indicates that the diffuse scattering that appears in planes at ($hk\pm\!\!\frac{l}{2}$) is mostly elastic within the energy resolution of the analyser.

\section{Conclusions}

We have probed the low-temperature magnetic structure of \sho\ by measuring the neutron diffraction in three planes orthogonal to the principal axes of the crystal. 
For this compound there appear to be two distinct types of short-range magnetic order, which can be inferred from the diffraction patterns, arising from the two crystallographically inequivalent sites for the Ho$^{3+}$ ions in the unit cell.
In the ($hk0$) plane diffuse magnetic scattering appears around the {\bf k}~=~0 positions at temperatures below 0.7~K. 
The Ho$^{3+}$ spins that participate in this short-range structure are antiferromagnetically coupled and are collinear with the $c$ axis.
Whereas another type of magnetic order forms nearly perfect planes of diffuse scattering which appear as ``rods'' of scattering intensity seen in both the ($h0l$) and ($0kl$) planes in reciprocal space at $Q$~=~(00$\frac{1}{2}$) and symmetry related positions.
This observation suggests that the second type of short-range order present in \sho\ is almost perfectly one-dimensional in nature. 
From the {\it XYZ}-polarized neutron measurements it is obvious that the spins that participate in this second structure are antiferromagnetically coupled and point along the $b$ axis.
The planes of diffuse scattering have internal structure, with maxima of intensity occurring at even integer $h$ positions at the lowest temperature.
Both types of diffuse scattering in \sho\ appear to be elastic within the energy resolution of the analyser.
 
The coexistence of two distinct types of magnetism is similar to what is observed at low-temperatures in \seo.
Both compounds have a magnetic component with the propagation vector {\bf k}~=~0 in the ($hk0$) plane below a well-defined transition temperature, as well as the one-dimensional scattering that is observed at much higher temperatures.

\section*{ACKNOWLEDGMENTS}

The authors would like to thank Mr.~T.~E.~Orton for the valuable technical support and acknowledge the dedication of the sample environment teams at both ISIS and ILL.
A part of this work was financially supported by the EPSRC, UK under grant EP/I007210/1.


\begin{thebibliography}{100}
\bibitem{2Gauss} No assumptions were made about the nature of the magnetic scattering distribution. The Lorentzian or Gaussian fitting functions were chosen because they gave the best fit to the data, assessed by comparing the value of $R^2$ and reduced $\chi^2$ for the different fits.
\bibitem{CorrCalc} Taking into account the instrument resolution by estimating the correletion lengths using $L$~=~$2\pi(W^2_{peak} - W^2_{ref})^{-\frac{1}{2}}$, where $W_{peak}$ and $W_{ref}$ are the full widths at half maximum in $Q$ space for the Lorentzian peak in question and a reference (resolution-limited) Bragg peak obtained on the same instrument resulted in a difference in the calculated correlation length of less than 1~$\%$ for the lengths along the $a$ and $b$ axes. 
The instrument resolution is more important for the calculations of correlation lengths along the $c$ axis where the scattering profile is much narrower.
\bibitem{CorrCalc2} Taking into account the instrument resolution did not result in a significant difference to the calculated correlation lengths.
\end{thebibliography}
\end{document}